\documentclass{article}
\usepackage[utf8]{inputenc}
\usepackage[margin=1in]{geometry}
\usepackage{slashed,graphicx}
\usepackage{amsthm}
\usepackage{amsmath}
\usepackage{float}
\usepackage{amssymb}
\usepackage[caption=false]{subfig}
\usepackage{hyperref}
\title{Study of $F$-wave Bottom Mesons in HQET}
\author{Ritu Garg, A. Upadhyay
 \\\small{\it School of Physics and Material Science},\\\small{\it Thapar University,
Patiala, Punjab-147004}\\\small{E-mail: ritugarg039@gmail.com,
alka.iisc@gmail.com}}

\begin{document}

\maketitle
\section{Abstract} We studied $F$-wave bottom mesons in heavy quark effective theory. The available experimental and theoretical data is used to calculate the masses of $F$-wave bottom mesons. The decay widths of bottom mesons are analyzed to find upper bounds of the associated couplings. We also construct Regge trajectories for our predicted data in planes ($J$, $M^2$ ) and our results nicely fit on Regge lines. Our results may provide a crucial information for upcoming experimental studies.

\section{Introduction}
In last 15 years, heavy-light hadrons are explored experimentally as well as theoretically. In D-meson family, several new states are observed by different experimental facilities like LHCb, BaBar, BESIII etc. which enriched charm spectrum. In 2021, LHCb observed state $D_{s0}^{0}(2590)$ with mass $M = 2591\pm 13$  MeV and decay width $\Gamma =  89\pm 28$ MeV respectively \cite{1}. They also assigned the state $D_{s0}^{0}(2590)$ with quantum number $n = 3$ and $l = 0$. Earlier in 2010, many candidates like $D(2550)^{0}$, $D^*(2600)^{0,+}$, $D(2750)^{0}$, $D^{*}(2760)^{0}$ were observed by BaBar collaborations \cite{2} and reconfirmed by LHCb in 2013 \cite{3}. Furthermore, LHCb collaborations in 2019 analysed four body amplitude of decay $B^{-} \longrightarrow D^{*+}\pi^{-}\pi^{+}$ \cite{4}. They reported existence of charm resonances $D_0(2550)^0$, $D_1^*(2600)^0$, $D_2(2740)^0$, $D_3^*(2760)$ with $J^P = 0^-, 1^-, 2^-$ and $3^-$ respectively. In addition, LHCb(2013) also confirmed states $D_J(3000)^{0}$ and $D_J^{*}(3000)^{0,+}$, with unnatural and natural parity respectively \cite{3}. In 2015, LHCb collaborations discovered the state $D_1^*(2760)^0$ with $J^P = 1^-$  by examine  $B^{-} \longrightarrow D^{+}\pi^{-}K^{-}$ decay \cite{5}.  However, bottom meson family is still less abundant in experimental confirmations. The only ground state $B^{0,\pm}(5279)$, $B^{*}(5324)$, $B_{s}(5366)$, $B_{s}^{*}(5415)$ and few orbitally excited states $B_{1}(5721)$, $B_{2}^{*}(5747)$ are listed in PDG \cite{6}. In 2013, CDF collaboration observed two higher resonances $B_J(5970)^{0,+}$ by analysing invariant mass distribution of $B^0\pi^+$ and $B^-\pi^+$ \cite{7}. In 2015, LHCb observed four resonances $B_1(5721)^{0,+}, B_2^*(5747), B_J(5840)^{0,+}, B_J(5960)^{0,+}$  by analysing mass spectra $B^+\pi^-, B^+\pi^- $ in p-p collision \cite{8}. The properties of state $B_J(5960)^{0,+}$ \cite{8} is consistent with the state $B_J(5970)^{0,+}$ of CDF(2013) \cite{7}. In $B_s$ family, ground state and 1P($1^+,2^+$) are well established. Recently (2021), LHCb detected the state $B_s^0$ of D-wave with two peaks of mass $6061\pm1.2\pm0.8$ MeV and $6114\pm3\pm5$ MeV in $B^+K^-$ mass spectrum \cite{9}. Apart from this, there is no experimental information for higher excited bottom meson states upto now. \\\\
In theory, various theoretical studies like masses, strong decays, radiative decays, weak decays, spin- parity value($J^P$) have done for higher excited bottom mesons with different models [10-37]. In B-meson family, the state $B^{0}(5279)$, $B^{\pm}(5279)$, $B^{*}(5324)$, $B_{s}(5366)$, $B_{s}^{*}(5416)$ are well established and classified as $1S$ state. In addition the state $B_{1}(5721)$, $B_{2}^{*}(5747)$ are experimentally confirmed and identified as $1P(1^+, 2^+)$ respectively. But there are puzzles for placing newly observed states $B_J(5840)^{0,+}, B_J(5960)^{0,+}$ in spectra. In literature , different theoretical models give different assignment for these states based upon predicted masses and decay widths reported by CDF and LHCb experimental groups. The newly observed state $B_J(5840)$ analyzed with quark model and favored the assignment of $B(2^1S_0)$ \cite{23,24}. But $^3 P_0$ decay model analysis suggested the assignments of state $B_J(5840)$ as $B(2^3S_1)$ \cite{27}. While Heavy quark effective theory explained resonances $B_J(5840)$ as $B(1^3D_1)$ state \cite{36}. The state $B_J(5960)^{0,+}$ can be assigned to $B(2^3S_1)$ with relativistic quark model \cite{11,21}. While some other models suggested it to be second orbitally excited $1^3D_3$ \cite{23,27} state or $1^3D_1$ state \cite{24}. There is also ambiguity for recently observed strange bottom meson states $B_{sJ}(6064)$ and $B_{sJ}(6114)$. Bing Chen et al. \cite{38} studied the states $B_{sJ}(6064)$ and $B_{sJ}(6114)$ with a non-relativistic quark potential model and suggested to be D-wave states.  Theoretically studies for these states are limited for-now which indicates it needs more attentions. The experimental progress stimulates the interest of theorists to check the validity  of the theoretically available models for these upcoming data. It also motivates us to explore the bottom mass spectrum theoretically for the missing states and fill the voids in spectrum.
\\ In this paper we studied the properties of $1F$ state by exploring Heavy quark effective theory. HQET is effective theory describes dynamics of heavy-light hadrons. In this theory, two kinds of approximate symmetries are incorporated: Heavy quark symmetry (HQS) and Chiral Symmetry of light quarks. The detailed information for HQET is discussed in section II (Theoretical framework).
With recent coming data, we are motivated to study 1$F$ states properties. In this paper, we predict masses and upper bounds of the associated couplings for $1F$ state. We analyzed decay behaviour of 1$F$ bottom meson states with pseudoscalar mesons and calculated upper bound on the associated couplings. This paper is summarized as: In section II, HQET and its lagrangian are briefly discussed. Using available data, masses and decays properties are studied in section III. In section IV, the conclusions drawn from our study are provided.
\section{Theoretical Formulation}
The study of heavy-light hadrons can be explored in framework HQET. HQET is powerful tool to describe properties of heavy-light mesons like masses, decay widths, branching ratios, fractions, spin, parity etc. This theory is explained with two approximate symmetries- heavy quark symmetry and chiral symmetry. Heavy quark symmetry is valid in approximation $m_Q\to\infty$. In limit $m_Q\rightarrow \infty$, spin of light quarks decoupled from spin of heavy quark, so total angular momentum of light quarks remains conserved. Total angular momentum of light quarks is $s_{l} = s_{q} + l$, $s_{q}$ = spin of light quark (1/2) and $l$ = total orbital momentum of light quarks. In heavy quark limit, mesons are grouped in doublets on basis of total angular momentum $s_{l}$ of light quarks. For $l = 0$, $s_{l} = 1/2$ coupled this with spin of heavy quark $s_{Q}$ = 1/2 and resulted with doublet $(0^{-},1^{-})$. This doublets is given by $(P,P^{*})$. For $l = 1$ forming two doublets denoted by $(P^{*}_{0}, P_{1}^{'})$ and $(P_{1}, P_{2}^{*})$ with $J_{s_{l}}^{P} = (0^{+}, 1^{+})$ and $J_{s_{l}}^{P} = (1^{+}, 2^{+})$ respectively. For $l = 2$, two doublets expressed by $(P^{*}_{1}, P_{2})$ and $(P_{2}^{'},P_{3}^{*})$ having $J_{s_{l}}^{P} = (1^{-}, 2^{-})$ and $J_{s_{l}}^{P} = (2^{-}, 3^{-})$ respectively. Similarly for $l = 3$, we get doublets $(P^{*}_{2}, P_{3})$ and $(P_{3}^{'},P_{4}^{*})$ for $J_{s_{l}}^{P} = (1^{-}, 2^{-})$ and $J_{s_{l}}^{P} = (2^{-}, 3^{-})$ respectively. These doublets are expressed in terms of super effective fields $H_{a}, S_{a}, T_{a}, X^{\mu}_{a}, Y^{\mu\nu}_{a}, Z^{\mu\nu}_{a}, R^{\mu\nu\rho}_{a}$ \cite{39} and expression for fields are given below:
\begin{gather}
\label{eq:lagrangian1}
 H_{a}=\frac{1+\slashed
v}{2}\{P^{*}_{a\mu}\gamma^{\mu}-P_{a}\gamma_{5}\}\\
S_{a} =\frac{1+\slashed v}{2}[{P^{'\mu}_{1a}\gamma_{\mu}\gamma_{5}}-{P_{0a}^{*}}]\\
T^{\mu}_{a}=\frac{1+\slashed v}{2}
\{P^{*\mu\nu}_{2a}\gamma_{\nu}-P_{1a\nu}\sqrt{\frac{3}{2}}\gamma_{5}
[g^{\mu\nu}-\frac{\gamma^{\nu}(\gamma^{\mu}-\upsilon^{\mu})}{3}]\}
\end{gather}
\begin{gather}
X^{\mu}_{a}=\frac{1+\slashed
v}{2}\{P^{\mu\nu}_{2a}\gamma_{5}\gamma_{\nu}-P^{*}_{1a\nu}\sqrt{\frac{3}{2}}[g^{\mu\nu}-\frac{\gamma_{\nu}(\gamma^{\mu}+v^{\mu})}{3}]\}
\end{gather}

\begin{multline}
 Y^{\mu\nu}_{a}=\frac{1+\slashed
v}{2}\{P^{*\mu\nu\sigma}_{3a}\gamma_{\sigma}-P^{'\alpha\beta}_{2a}\sqrt{\frac{5}{3}}\gamma_{5}[g^{\mu}_{\alpha}g^{\nu}_{\beta}
-\frac{g^{\nu}_{\beta}\gamma_{\alpha}(\gamma^{\mu}-v^{\mu})}{5}-\frac{g^{\mu}_{\alpha}\gamma_{\beta}(\gamma^{\nu}-v^{\nu})}{5}]\}   
\end{multline}
\begin{multline}
Z^{\mu\nu}_{a}=\frac{1+\slashed
v}{2}\{P^{\mu\nu\sigma}_{3a}\gamma_{5}\gamma_{\sigma}-P^{*\alpha\beta}_{2a}\sqrt{\frac{5}{3}}[g^{\mu}_{\alpha}g^{\nu}_{\beta}
-\frac{g^{\nu}_{\beta}\gamma_{\alpha}(\gamma^{\mu}+v^{\mu})}{5}-\frac{g^{\mu}_{\alpha}\gamma_{\beta}(\gamma^{\nu}+v^{\nu})}{5}]\}
\end{multline}
\begin{multline}
R^{\mu\nu\rho}_{a}=\frac{1+\slashed
v}{2}\{P^{*\mu\nu\rho\sigma}_{4a}\gamma_{5}\gamma_{\sigma}-P^{'\alpha\beta\tau}_{3a}\sqrt{\frac{7}{4}}[g^{\mu}_{\alpha}g^{\nu}_{\beta}g^{\rho}_{\tau}
-\frac{g^{\nu}_{\beta}g^{\rho}_{\tau}\gamma_{\alpha}(\gamma^{\mu}-v^{\mu})}{7}-\frac{g^{\mu}_{\alpha}g^{\rho}_{\tau}\gamma_{\beta}(\gamma^{\nu}-v^{\nu})}{7}-\frac{g^{\mu}_{\alpha}g^{\nu}_{\beta}\gamma_{\tau}(\gamma^{\rho}-v^{\rho})}{7}]\}
\end{multline}
The field $H_a$ shows doublets of $S$-wave for $J^P = (0^-,1^-)$. The fields $S_a$ and $T_a$ describes doublets of $P$-wave for $J^P = (0^+, 1^+)$ and $(1^+, 2^+)$ respectively. $D$-wave doublets for $J^P = (1^-, 2^-)$ and $(2^-, 3^-)$ belongs to fields $X^{\mu}_{a}$ and $Y^{\mu\nu}_{a}$ respectively. In same manner, fields $Z^{\mu\nu}_{a}$, $R^{\mu\nu\rho}_{a}$ presents doublets of $F$-wave for $J^P = (2^+, 3^+)$ and $(3^+, 4^+)$ respectively. $a$ in above expressions is light quark ($u, d, s$) flavor index. $v$ is heavy quark velocity, conserved in strong interactions. The approximate chiral symmetry $SU(3)_L\times SU(3)_R$ is involved with fields of pseudoscalar mesons $\pi$, K, and $\eta$ which are lightest strongly interacting  bosons. They are treated as approximate Goldstone bosons of this chiral symmetry and can be introduced by the matrix field $U(x) = Exp\left[\dot{\iota}\sqrt{2}\phi(x)/f\right]$, where $\phi(x)$ is given by
\begin{center}
\begin{equation}
\mathcal{M} = \begin{pmatrix}
\frac{1}{\sqrt{2}}\pi^{0}+\frac{1}{\sqrt{6}}\eta & \pi^{+} & K^{+}\\
\pi^{-} & -\frac{1}{\sqrt{2}}\pi^{0}+\frac{1}{\sqrt{6}}\eta &
K^{0}\\
K^{-} & \overline{K}^{0} & -\sqrt{\frac{2}{3}}\eta
\end{pmatrix}
\end{equation}
\end{center}
The fields of heavy meson doublets is given in eqn.(1-7) interact with pseudoscalar goldstone bosons via covariant derivative $D_{\mu ab}= -\delta_{ab}\partial_{\mu}+\mathcal{V}_{\mu ab} =  -\delta_{ab}\partial_{\mu}+\frac{1}{2}(\xi^{+}\partial_{\mu}\xi+\xi\partial_{\mu}\xi^{+})_{ab}$ and axial vector field $A_{\mu ab}=\frac{i}{2}(\xi\partial_{\mu}\xi^{\dag}-\xi^{\dag}\partial_{\mu}\xi)_{ab}$. By including all meson doublet fields and goldstone fields, effective lagranigan is given by:
\begin{multline}
\mathcal{L} = iTr[\overline{H}_{b}v^{\mu}D_{\mu ba}H_{a}]+ \frac{f_\pi^{2}}{8}Tr[\partial^{\mu}\Sigma\partial_{\mu}\Sigma^{+}] + Tr[\overline{S_{b}}(iv^{\mu}D_{\mu ba} - \delta_{ba}\Delta_{S})S_{a}]+Tr[\overline{T_{b}^{\alpha}}(iv^{\mu}D_{\mu ba}- \delta_{ba}\Delta_{T})T_{a \alpha}\\+ Tr[\overline{X_{b}^{\alpha}}(iv^{\mu}D_{\mu ba}- \delta_{ba}\Delta_{X})X_{a \alpha}+ Tr[\overline{Y_{b}^{\alpha\beta}}(iv^{\mu}D_{\mu ba}- \delta_{ba}\Delta_{Y})Y_{a\alpha\beta}+Tr[\overline{R_{b}^{\alpha\beta\rho}}(iv^{\mu}D_{\mu ba}- \delta_{ba}\Delta_{R})T_{a \alpha\beta\rho}] 
\end{multline}
The mass parameter $\Delta_{F}$ in equation (9) presents the mass difference between higher mass doublets ($F$) and lowest lying doublet ($H$) in terms of spin average masses of these doublets with same principle quantum number (n). The expressions for mass parameters are given by:
\begin{align}
             \Delta_{F}=\overline{M_{F}}&- \overline{M_{H}},~~ F= S,T,X,Y,Z,R\\
\text{where, }~~~~~~~~~~~
           \overline{M_{H}}&=(3m^{Q}_{P_1^*}+m^{Q}_{P_{0}})/4\\
         \overline{M_{S}}&=(3m^{Q}_{P_1^{'}}+m^{Q}_{P_0^*})/4\\
        \overline{M_{T}}&=(5m^{Q}_{P_2^*}+3m^{Q}_{P_1})/8\\
          \overline{M_{X}}&=(5m^{Q}_{P_2}+3m^{Q}_{P_1^{*}})/8\\
           \overline{M_{Y}}&=(5m^{Q}_{P_3^*}+3m^{Q}_{P_2^{'}})/8\\
            \overline{M_{Z}}&=(7m^{Q}_{P_3}+5m^{Q}_{P_2^*})/12\\
             \overline{M_{R}}&=(9m^{Q}_{P_4^*}+7m^{Q}_{P_{3}^{'}})/12
         \end{align}
           The $1/m_{Q}$ corrections to heavy quark limit are given by symmetry breaking terms. The corrections are form of: 
           \begin{multline}
          \mathcal{L}_{1/m_{Q}} = \frac{1}{2m_{Q}}[\lambda_{H} Tr(\overline H_{a}\sigma^{\mu\nu}{H_{a}}\sigma_{\mu\nu}) + \lambda_{S}Tr(\overline S_{a}\sigma^{\mu\nu} S_{a}\sigma_{\mu\nu})+\lambda_{T}Tr(\overline T_{a}^{\alpha}\sigma^{\mu\nu}{T_{a}^{\alpha}}\sigma_{\mu\nu})+\lambda_{X}Tr(\overline X_{a}^{\alpha}\sigma^{\mu\nu}{X_{a}^{\alpha}}\sigma_{\mu\nu})\\ + \lambda_{Y}Tr(\overline Y_{a}^{\alpha\beta}\sigma^{\mu\nu}{Y_{a}^{\alpha\beta}}\sigma_{\mu\nu}) +\lambda_{Z}Tr(\overline Z_{a}^{\alpha\beta}\sigma^{\mu\nu}{Z_{a}^{\alpha\beta}}\sigma_{\mu\nu})+ \lambda_{R}Tr(\overline R_{a}^{\alpha\beta\rho}\sigma^{\mu\nu}{R_{a}^{\alpha\beta\rho}}\sigma_{\mu\nu})]
          \end{multline} 
          Here parameters $\lambda_{H}$, $\lambda_{S}$, $\lambda_{T}$, $\lambda_{X}$, $\lambda_{Y}$, $\lambda_{Z}$, $\lambda_{R}$ are analogous with hyperfine splittings and defined as in Eq.(19 - 25). This mass terms in lagrangian represent only first order in $1/m_{Q}$ terms, but higher order terms may also be present otherwise. We are restricting to the first order corrections in $1/m_{Q}$.
 \begin{align}
     \lambda_{H} = \frac{1}{8}(M^{2}_{P^{*}} - M^{2}_{P}) \\
     \lambda_{S} = \frac{1}{8}({M^{2}_{P_1^{'}}} - {M^{2}_{P_0^*}})\\
     \lambda_{T}=\frac{3}{16}({M^{2}_{P_2^*}}-{M^{2}_{P_1}})\\
     \lambda_{X}=\frac{3}{16}({M^{2}_{P_2}}-{M^{2}_{P_1^*}})\\
      \lambda_{Y}=\frac{5}{24}({M^{2}_{P_3}}-{M^{2}_{P_2^{'*}}})\\
      \lambda_{Z}=\frac{5}{24}({M^{2}_{P_3^*}}-{M^{2}_{P_2^{'}}})\\
      \lambda_{R}=\frac{7}{32}({M^{2}_{P_4^*}}-{M^{2}_{P_3^{'*}}})
 \end{align}

   In HQET, at scale of 1 GeV flavor symmetry spontaneously arises for b (bottom quark) and c (charm quark) and hence elegance of flavor symmetry implies to
        \begin{align}
           \label{1eu_eqn}
           \Delta_{F}^{(c)} =\Delta_{F}^{(b)}\\
   \lambda_{F}^{(c)} = \lambda_{F}^{(b)}
   \label{2eu_eqn}
\end{align}

The decays $F\rightarrow  H + M$ (F = $H, S, T, X, Y, Z, R$
and $M$ presents a light pseudoscalar meson) can be described by effective Lagrangians explained in terms of the fields introduced in (9-14) that valid at leading order in the heavy quark mass and light meson momentum expansion:
\begin{gather}
\label{eq:lagrangian17}
L_{HH}=g_{HH}Tr\{\overline{H}_{a}
H_{b}\gamma_{\mu}\gamma_{5}A^{\mu}_{ba}\}\\
L_{TH}=\frac{g_{TH}}{\Lambda}Tr\{\overline{H}_{a}T^{\mu}_{b}(iD_{\mu}\slashed
A + i\slashed D A_{\mu})_{ba}\gamma_{5}\}+h.c.\\
L_{XH}=\frac{g_{XH}}{\Lambda}Tr\{\overline{H}_{a}X^{\mu}_{b}(iD_{\mu}\slashed
A + i\slashed D A_{\mu})_{ba}\gamma_{5}\}+h.c.
\end{gather}
\begin{multline}
L_{YH}=\frac{1}{\Lambda^{2}}Tr\{\overline{H}_{a}Y^{\mu\nu}_{b}[k^{Y}_{1}\{D_{\mu}
,D_{\nu}\}A_{\lambda}+k^{Y}_{2}(D_{\mu}D_{\lambda}A_{\nu}
+D_{\nu}D_{\lambda}A_{\mu})]_{ba}\gamma^{\lambda}\gamma_{5}\}+h.c.
\end{multline}
\begin{multline}
L_{ZH}=\frac{1}{\Lambda^{2}}Tr\{\overline{H}_{a}Z^{\mu\nu}_{b}[k^{Z}_{1}\{D_{\mu}
,D_{\nu}\}A_{\lambda}+
k^{Z}_{2}(D_{\mu}D_{\lambda}A_{\nu}+D_{\nu}D_{\lambda}A_{\mu})]_{ba}\gamma^{\lambda}\gamma_{5}\}+h.c.
\end{multline}
\begin{multline}
L_{RH}=\frac{1}{\Lambda^{3}}Tr\{\overline{H}_{a}R^{\mu\nu\rho}_{b}[k^{R}_{1}\{D_{\mu}
,D_{\nu}D_{\rho}\}A_{\lambda}+k^{R}_{2}(\{D_{\mu},D_{\rho}\}D_{\lambda}A_{\nu}+\{D_{\nu},D_{\rho}\}D_{\lambda}A_{\mu}
+\{D_{\mu},D_{\nu}\}D_{\lambda}A_{\rho})]_{ba}\gamma^{\lambda}\gamma_{5}\}\\+h.c.
\end{multline}
In these equations $D_{\mu} =
\partial_{\mu}+V_{\mu}$,  $\{D_{\mu},D_{\nu}\}
= D_{\mu}D_{\nu}+D_{\nu}D_{\mu}$ and $\{D_{\mu} ,D_{\nu}D_{\rho}\} =
D_{\mu}D_{\nu}D_{\rho}+D_{\mu}D_{\rho}D_{\nu}+D_{\nu}D_{\mu}D_{\rho}+D_{\nu}D_{\rho}D_{\mu}+D_{\rho}D_{\mu}
D_{\nu}+D_{\rho}D_{\nu}D_{\mu}$. $\Lambda$ is the chiral symmetry
breaking scale taken as 1 GeV. $g_{HH}$, $g_{SH}$, $g_{TH}$, $g_{YH}
= k^{Y}_{1}+k^{Y}_{2}$ and $g_{ZH} = k^{Z}_{1}+k^{Z}_{2}$ are the
strong coupling constants involved.  Using the lagrangians
$L_{HH}, L_{SH}, L_{TH}, L_{YH}, L_{ZH}, L_{RH}$, the two body strong decays of $Q\overline{q}$ heavy-light bottom mesons are given as
$(2^{+},3^{+}) \rightarrow (0^{-},1^{-}) + M$
\begin{gather}
\label{eq:lagrangian2} \Gamma(2^{+} \rightarrow 1^{-})=
C_{M}\frac{8g_{ZH}^{2}}{75\pi f_{\pi}^{2}\Lambda^{4}}
\frac{M_{f}}{M_{i}}[p_{M}^{5}(m_{M}^{2}+p_{M}^{2})]\\
\Gamma(2^{+} \rightarrow 0^{-})= C_{M}\frac{4g_{ZH}^{2}}{25\pi f_{\pi}^{2}\Lambda^{4}}\frac{M_{f}}{M_{i}}[p_{M}^{5}(m_{M}^{2}+p_{M}^{2})]\\
\Gamma(3^{+} \rightarrow 1^{-})= C_{M}\frac{4g_{ZH}^{2}}{15\pi
f_{\pi}^{2}\Lambda^{4}}
\frac{M_{f}}{M_{i}}[p_{M}^{5}(m_{M}^{2}+p_{M}^{2})]
\end{gather}
$(3^{+},4^{+}) \rightarrow (0^{-},1^{-}) + M$
\begin{gather}
\label{eq:lagrangian} \Gamma(3^{+} \rightarrow 1^{-})=
C_{M}\frac{36g_{RH}^{2}}{35\pi f_{\pi}^{2}\Lambda^{6}}
\frac{M_{f}}{M_{i}}[p_{M}^{9}]\\
\Gamma(4^{+} \rightarrow 1^{-})= C_{M}\frac{4g_{RH}^{2}}{7\pi f_{\pi}^{2}\Lambda^{6}}\frac{M_{f}}{M_{i}}[p_{M}^{9}]\\
\Gamma(4^{+} \rightarrow 0^{-})= C_{M}\frac{16g_{RH}^{2}}{35\pi
f_{\pi}^{2}\Lambda^{6}}
\frac{M_{f}}{M_{i}}[p_{M}^{9}]
\end{gather}
where $M_{i}$, $M_{f}$ represents initial and final momentum, $\Lambda$ is chiral symmetry breaking scale of 1 GeV. $p_{M}$, $m_{M}$ denotes to final momentum and mass of light pseudoscalar meson. Coupling constant plays key role in phenomenology study of heavy light mesons. These dimensionless coupling constants describes strength of transition between H-H field (negative-negative parity), S-H field (positive-negative parity), T-H field (positive-negative parity).
The coefficient $C_{M}$ for different pseudoscalar particles are:
$C_{\pi^{\pm}}$, $C_{K^{\pm}}$, $C_{K^{0}}$, $C_{\overline{K}^{0}}=1$, $C_{\pi^{0}}=\frac{1}{2}$ and $C_{\eta}=\frac{2}{3}(c\bar{u}, c\bar{d})$ or $\frac{1}{6}(c\bar{s})$. In our paper, we are not including higher order corrections of $\frac{1}{m_{Q}}$ to bring new couplings. We also expect that higher corrections give small contribution in comparison of leading order contributions.
\section{Numerical Analysis:}
Recently observed states like $D_{0}(2560)$, $D_{1}^{*}(2680)$,
$D_{2}(2740)$, $D_{3}^{*}(2760)$, $D_{J}(3000)$, $D_{J}^{*}(3000)$ and strange states $D_{s1}(2860)$, $D_{s3}(2860)$, $D_{s}(3040)$ have stimulated charm sector. But in case of bottom sector, there is less experimental states as compare to charm sector. Newly observed excited strange bottom meson states $B_{sJ}(6064)$ and $B_{sJ}(6114)$ have developed interest of theoreticians to study excited states of bottom sector. With coming data from different experimental facilities, we are motivated to predict masses and upper bound of coupling constants for $1F$ bottom meson states with strange partners in framework of HQET.\\
In this paper, our calculations are divided into two parts- first is to calculate masses of $1F$ bottom states and in second , we have studied decay behaviour of these states.\\
\subsection{Masses}
To describe spectroscopy of bottom and bottom-strange mesons, mass is one of an important parameter. Input values used for calculating masses of $1F$ bottom states are listed in Table I.
\begin{table}[ht]
        \centering
       \caption{Input values used in this work. All values are in units of MeV.}
          \begin{tabular}{|c|c|c|c|c|c|}
      \hline
     State & $J^{P}$ & $ c\overline{q}$ & $ c\overline{s}$ & $ b\overline{q}$ & $ b\overline{s}$ \\
     \hline
     $1^{1}S_{0}$ & $0^{-}$ & 1869.50 \cite{6}& 1969.0 \cite{6} & 5279.5 \cite{6} & 5366.91 \cite{6}  \\
     \hline
      $1^{3}S_{1}$ & $1^{-}$ & 2010.26 \cite{6} & 2106.6 \cite{6} & 5324.71 \cite{6} & 5415.8 \cite{6} \\
      \hline
      $1^{3}F_{2}$ & $2^{+}$ & 3132 \cite{41} & 3208 \cite{41} & - & -  \\
      \hline
      $1F_{3}$ & $3^{+}$ & 3143 \cite{41} & 3218 \cite{41} & - & -  \\
      \hline
      $1F^{'}_{3}$ & $3^{+}$ &3108 \cite{41}  & 3186 \cite{41} & - & -  \\
      \hline
      $1^{3}F_{4}$ & $2^{+}$ & 3113 \cite{41}  & 3190 \cite{41}& - & -  \\
      \hline

     \end{tabular}
    \label{tab:my_label1}
\end{table}

 To compute the masses of $1F$  bottom states, primarily we calculated value of average masses $\overline{M_{H}}$, $\overline{M_{Z}}$, $\overline{M_{R}}$ introduced in eqn (10-17) for charm mesons states from Table I, then heavy quark symmetry (HQS) parameters $\Delta_{Z}$, $\Delta_{R}$, $\lambda_{Z}$, $\lambda_{R}$  described in eqn (10, 24, 25) are calculated for same charm meson states. The parameters $\Delta_{F}$, $\lambda_{F}$ are flavor independent in HQET. It implies $\Delta_{F}^{(c)} = \Delta_{F}^{(b)}$,  $\lambda_{F}^{(c)} = \lambda_{F}^{(b)}$.
With the help of calculated flavor symmetry parameters $\Delta_{F}$,  $\lambda_{F}$ for charm mesons and then applying heavy quark symmetry,  we predicted masses for $1F$ bottom mesons listed in Table II. For details, we are elaborating calculation part of mass of $B(1^{3}F_{2})$. From Table I, using the charm states we calculated $\overline{M^c_{H}}$ = 1975.07 MeV, $\overline{M^c_{Z}}$ = 3138.42 MeV. Then using these two values, we have $\Delta^{(c)}_{Z}=\overline{M^c_{Z}}-\overline{M^c_{H}}= 1163.35$ MeV. Using eqn. (24) we get $\lambda^{(c)}_Z=14380.2$ MeV$^2$. The symmetry of these parameters are given by eqn. (26) and (27) implies that $\Delta^{(b)}_{Z}= 1163.35$ MeV and $\lambda^{(b)}_Z=14380.2$ MeV$^2$. Similarly from Table I, we calculated $\overline{M^b_{H}} = 5313.36$ MeV for bottom mesons. Using values of $\Delta^{(b)}_{Z}= 1163.35$ MeV $\lambda^{(b)}_Z=14380.2$ MeV$^2$ and $\overline{M^b_{H}} = 5313.36$ MeV, we obtained masses of $1F$ bottom mesons listed in Table II. For comparison, predictions from different models are also mentioned in Table II. The masses obtained using the heavy quark symmetry in our work are in agreement with the masses obtained by the relativistic quark model in Ref.\cite{10} with deviation of $\pm 1.2\%$ for non strange states while  strange states are deviated by $\pm 0.6\%$. On comparing with Ref.\cite{22}, our results are deviated by $\pm 3\%$. So, our results are in overall good agreement with other  theoretical models. We have also explored symmetry parameters ($\Delta_{F}$
   $\lambda_{F}$) by taking different mass sets from Table II. The computed values of parameters are listed in Table III. The parameter $\Delta_{Z}$, $\Delta_{Z}$ are consistent for different predicted theoretical masses from Table II. The parameters $\lambda_{Z}$, $\lambda_{Z}$ are also close to each other for same sets of masses. As $\lambda_{Z}$, $\lambda_{Z}$ are responsible for hyperfine splittings of Z and R fields respectively, we find the masses of $1F$ states with above parameters are in reasonable agreement with other theoretical estimates.   
\begin{table}
\centering
\caption{Obtained masses for $1F$ bottom mesons}
 \begin{tabular}{| c | c | c | c | c | c | c |}
      \hline
      \multicolumn{1}{|c|}{} & \multicolumn{6}{c|}{Masses of $1F$ Bottom Mesons (MeV)}\\
       \cline{2-7}
     
       \multicolumn{1}{|c|}{$J^{P}$}&\multicolumn{3}{c|}{Non-Strange}&\multicolumn{3}{c|}{Strange}\\
       \cline{2-7}
       & \multicolumn{1}{c|}{Calculated}&\multicolumn{1}{c|}{\cite{10}}&\multicolumn{1}{c|}{\cite{22}}&\multicolumn{1}{c|}{Calculated}&\multicolumn{1}{c|}{[10]}&\multicolumn{1}{c|}{[23]}\\
       \hline
        $2^{+}(1^3F_2)$ &  6473.6 & 6412 &  6387& 6518.28 & 6501 & 6358\\
        $3^{+}(1F_3)$ &  6478.93 & 6420 & 6396 & 6523.21 & 6515 & 6369\\
          $3^{+}(1F^{'}_3)$ &  6447.76 & 6391 & 6358 & 6506.05 & 6468 & 6318\\
          $4^{+}(1^3F_4)$ &  6450.14& 6380 & 6364 & 6508.01 & 6475& 6328\\
      \hline
   \end{tabular}
   \end{table}
   \\
   \begin{table}[htp]
    \centering
    \caption{ Values of symmetry parameters.}
    \begin{tabular}{|c|c|c|c|c|}
    \hline
       Parameters  & Our Calculations & Ref.[10] & Ref.[23]\\
         \hline
         $\Delta_{Z}$ (MeV) & 1163.35 & 1103.31 & 1078.89\\
         \hline
         $\Delta_{R}$ (MeV) &  1135.74 & 1071.45& 1048.02\\
         \hline
         $\lambda_{Z}$(GeV$^2$) & 0.014 & 0.021& 0.023\\
         \hline
          $\lambda_{R}$(GeV$^2$)& 0.007 & -0.03& 0.017\\
         \hline
        \end{tabular}
  
    \label{tab:my_label62}
\end{table}
\subsection{Decay Widths}
    By using obtained masses, we computed decay width for $1F$ bottom mesons via pseudoscalar particles ($\pi,\eta,K$) only in form of coupling constants. Formulation for decay widths are discussed in Section II. These decay width formulas are for strong decays via pseudoscalar particles only. Input values used for calculating decay width are $M_{\pi^{0}}$ = 134.97 MeV, $M_{\pi^{+}}$ = 139.57 MeV, $ M_{K^{+}}$= 493.67 MeV, $M_{\eta^{0}}$= 547.85 MeV, $ M_{K^{0}}$= 497.61 MeV, $M_{B^{*\pm}} = 5324.70\pm 0.12$ MeV, $M_{B^{\mp}} = 5279.34\pm0.12$ MeV, $M_{B^{0}_{S}} = 5366.88\pm0.14$ MeV   $M_{B^{*\pm}_{S}}= 5415.40$ MeV and calculated masses for $F$-wave, bottom mesons listed in Table II.\\
 The computed strong decay widths in form of coupling constants $g_{ZH}$, $g_{RH}$ for $1F$ bottom mesons are collected in Table III and Table IV. The computed decay widths for bottom meson states does not including weak decays as well as radiative decays and also excludes decays via emissions of vector mesons ($\omega,\rho,K^*,\phi$). So, on comparing these computed decay widths with theoretical available total decay widths \cite{41} give upper bounds of associated couplings ($g_{ZH}$, $g_{RH}$). Coupling constant plays important role in hadron spectroscopy. Here dimensionless coupling constants $g_{ZH}$, $g_{RH}$ give strength of transitions between Z-H fields and R-H fields respectively. Values of coupling constants are more for ground state transitions (H-H fields) than excited states (S-H, T-H, X-H, Z-H, R-H fields) transition shown by value of $g_{HH} =0.64\pm0.075$ \cite{39} while $g_{SH} =0.56\pm0.04$, $g_{TH} =0.43\pm0.01$ \cite{39}, $g_{XH}$ = 0.24 \cite{32}. Also values of coupling  constants are low at higher orders (n = 2, n = 3) in comparison to lower order (n = 1) interactions \cite{39,43} like $\tilde{g}_{HH} = 0.28\pm 0.015$, $\tilde{g}_{sH} = 0.18\pm 0.01$ so on. This progression also supports the values of coupling computed here. A particular state like B(6473) give 22538.78$g_{ZH}^2$ total decay width, when compared with total decay widths calculated by other theoretical paper \cite{41}, we provided an upper bound on $g_{ZH}$ value. Now if we add additional modes, then value of $g_{ZH}$ may be lesser than 0.09 ($g_{ZH}< 0.09$). So, these upper bounds may give important information to other associated bottom states. Without enough experimental information, it is not possible to compute values of coupling constants from heavy quark symmetry entirely but upper bound to these coupling are mentioned in Table IV and Table V. In our study, we are taking  decays modes via pseduoscalar mesons only.
  Here, we need to emphasize that computed total decay widths for above states do not include contribution of decays via emission of vector mesons ($\omega,\rho,K^*,\phi$). Since the contribution of vector mesons to total decay widths are small as compare to pseudoscalar mesons. They give contribution of $\pm50$ MeV to total decay widths for above analysed states\cite{22}. Further, we are discussed Regge Trajectories which justify our calculated masses of $1F$ bottom meson states.
   \begin{table*}{\normalsize
\renewcommand{\arraystretch}{1.0}
\tabcolsep 0.2cm \caption{\label{tab:expt} Decay Width of obtained masses for $1F$ bottom mesons.}
 \centering
   \begin{tabular}{|c|c|c|c|c|c|}
    \hline
    States & $J^{P}$ & Decay Modes& Decay Widths& Total Decay Width[MeV]\cite{41}& Upper bound of coupling constant \\
    \hline
    $B(6473.6)$ & $2^{+}$ & $B^{*+}\pi^-$ & 3347.69$g_{ZH}^2$&&\\
    & & $B^{*0}\pi^{0}$ & 1672.14$g_{ZH}^2$ &&\\
    & & $B^{*0}\eta^{0}$&1367.92$g_{ZH}^2$&&\\
    & & $B^{*}_{s}K^{0}$ & 1328.31$g_{ZH}^2$&&\\
     & & $B^{+}\pi^{-}$ & 6262.1$g_{ZH}^2$&&\\
      & & $B^{0}\eta^{0}$ & 2657.76$g_{ZH}^2$&&\\
       & & $B^{0}_{s}K^{0}$ & 2774.61$g_{ZH}^2$&&\\
        & & $B^{0}\pi^{0}$ & 3128.25$g_{ZH}^2$&&\\
    & &Total & 22538.78$g_{ZH}^2$ && \\
    & & & $g_{ZH}$& 202.4 & 0.09 \\
     \hline
     $B(6478.93)$ & $3^{+}$ & $B^{*0}\pi^0$ &  4609.40$g_{ZH}^2$&& \\
      & &  $B^{*+}\pi^-$ & 9201.01$g_{ZH}^2$&& \\
      & & $B^{*0}\eta^0$ & 3823.83$g_{ZH}^2$&& \\
       & & $B_{s}^{*0}K^0$ & 3700.1$g_{ZH}^2$&& \\
          & &Total & 21334.34$g_{ZH}^2$& &  \\
          & & & $g_{ZH}$&105.2 & 0.07 \\
    \hline
   $B(6447.76)$ & $3^{+}$ & $B^{*0}\pi^0$ &  29675$g_{RH}^2$&& \\
      & &  $B^{*+}\pi^-$ & 14898.6$g_{RH}^2$&& \\
      & & $B^{*0}\eta^0$ & 7117.82$g_{RH}^2$&& \\
       & & $B_{s}^{*0}K^0$ & 5933.87$g_{RH}^2$&& \\
          & &Total & 57625.29$g_{RH}^2$ &&  \\
          & & & $g_{RH}$ & 221.8& 0.06\\ 
     \hline
      $B(6450.14)$ & $4^{+}$ & $B^{*+}\pi^-$ & 16168$g_{RH}^2$&&\\
    & & $B^{*0}\pi^{0}$ & 8117.39$g_{RH}^2$ &&\\
    & & $B^{*0}\eta^{0}$&3856.74$g_{RH}^2$&&\\
    & & $B^{*}_{s}K^{0}$ & 3208.9$g_{RH}^2$&&\\
     & & $B^{+}\pi^{-}$ & 21811.5$g_{RH}^2$&&\\
     & & $B^{0}\eta^{0}$ & 5638.37$g_{RH}^2$&&\\
       & & $B^{0}_{s}K^{0}$ & 5731.86$g_{RH}^2$&&\\
        & & $B^{0}\pi^{0}$ & 10925.9$g_{ZH}^2$&&\\
    & & Total & 22538.78$g_{ZH}^2$ & &\\
    & & & $g_{RH}$& 110.0 & 0.07 \\
          \hline
    \end{tabular}
    }
    \end{table*}
   \begin{table*}{\normalsize
\renewcommand{\arraystretch}{1.0}
\tabcolsep 0.2cm \caption{\label{tab:expt1} Decay Width of obtained masses for $1F$ strange bottom mesons.}
 \centering
   \begin{tabular}{|c|c|c|c|c|c|}
    \hline
    States & $J^{P}$ & Decay Modes& Decay Widths& Total Decay width[MeV] \cite{41} & Upper bound of coupling constant \\
    \hline
    $B_{s}(6518.28)$ & $2^{+}$ & $B^{*0}K^0$ & 1850.44$g_{ZH}^2$&&\\
    & & $B^{*-}K^{+}$ & 1864.99$g_{ZH}^2$ &&\\
    & & $B_{s}^{*0}\eta^{0}$&152.47$g_{ZH}^2$&&\\
    & & $B^{*}_{s}\pi^{0}$ & 861.63$g_{ZH}^2$&&\\
     & & $B^{0}K^{0}$ & 3601.08$g_{ZH}^2$&&\\
      & & $B^{-}K^{+}$ & 3632.92$g_{ZH}^2$&&\\
       & & $B^{0}_{s}\eta^{0}$ & 317.73$g_{ZH}^2$&&\\
        & & $B_{s}^{0}\pi^{0}$ & 1687.81$g_{ZH}^2$&&\\
    & &Total & 13969.07$g_{ZH}^2$ && \\
    & & & $g_{ZH}$&256.3 & 0.13 \\
     \hline
     $B_{s}(6523.21)$ & $3^{+}$ & $B^{*-}K^+$ &  4762.7$g_{ZH}^2$&& \\
      & &  $B^{*0}K^0$ & 4727.53$g_{ZH}^2$&& \\
      & & $B_{s}^{*0}\eta^0$ & 394.77$g_{ZH}^2$&& \\
       & & $B_{s}^{*0}\pi^0$ & 2216.24$g_{ZH}^2$&& \\
          & &Total & 12101.24$g_{ZH}^2$ &&  \\
          & & & $g_{ZH}$&138.4 & 0.10 \\
    \hline
   $B_{s}(6506.05)$ & $3^{+}$ & $B^{*-}K^+$ &  12627.5$g_{RH}^2$&& \\
      & &  $B^{*0}K^0$ & 12434.2$g_{RH}^2$&& \\
      & & $B_{s}^{*0}\eta^0$ & 714.78$g_{RH}^2$&& \\
       & & $B_{s}^{*0}\pi^0$ & 7469.58$g_{RH}^2$&& \\
          & &Total & 33,246.06$g_{RH}^2$ & & \\
          & & & $g_{RH}$&274 & 0.09 \\
     \hline
       $B_{s}(6508.01)$ & $4^{+}$ & $B^{*0}K^0$&  7023.53$g_{RH}^2$&&\\
    & & $B^{*-}K^{+}$ & 7131.87$g_{RH}^2$ &&\\
    & & $B_{s}^{*0}\eta^{0}$&405.17$g_{RH}^2$&&\\
    & & $B^{*}_{s}\pi^{0}$ & 4211.83$g_{RH}^2$&&\\
     & & $B^{0}K^{0}$ & 8073.71$g_{RH}^2$&&\\
      & & $B^{-}K^{+}$ & 8207.52$g_{RH}^2$&&\\
       & & $B^{0}_{s}\eta^{0}$ &518.27 $g_{RH}^2$&&\\
        & & $B_{s}^{0}\pi^{0}$ & 4787.05$g_{RH}^2$&&\\
    & &Total & 40358.95$g_{RH}^2$ & & \\
    & & & $g_{RH}$&138.6 & 0.05 \\
          \hline
    \end{tabular}
    }
    \end{table*} 
    \subsection{Regge Trajectory}
    Regge trajectories are a effective phenomenological approach to describe hadrons spectroscopy. The plotted between total angular momentum($J$) and radial quantum number($n_r$) of hadrons against square of their masses($M^2$) provides information about quantum number of particular state and also helps to identify recently observed states. We use following definitions:
    \begin{itemize}
        \item[(a).]{The ($J$,$M^2$ ) Regge trajectories:
    \begin{equation}
        J = \alpha M^2 + \alpha_{0}
      \end{equation}}
      \item[(b).]{The ($n_r$, $M^2$ ) Regge trajectories:
    \begin{equation}
       n_r  = \beta M^2 + \beta_{0}
    \end{equation} }
    \end{itemize}
      here $\alpha$, $\beta$ are slopes and $\alpha_{0}$, $\beta_{0}$ are intercepts. We plot Regge trajectories in plane (J, $M^2$ ) with natural parity $P= (-1)^J$ and unnatural parity $P= (-1)^{J-1}$ for $1F$ bottom mesons using predicted spectroscopic data. The plots of Regge trajectories in the ($J,M^2$) plane are also known as Chew-frautschi plots  \cite{44,45,46}. The plots are shown in fig. $1-4$. In fig $1-4$, masses for S-wave and P-wave($1^+,2^+$) are taken from PDG; remaining masses are taken from Ref. [14] and for 1F, we are taking our calculated masses. As Regge trajectories are known to be linear for mesons,it supports our obtained results and also helps to define spin parity state to particular resonance mass. Since Regge lines are almost linear, parallel and equidistant in fig $1-4$. Regge slopes ($\alpha$) and Regge intercepts $\alpha_{0}$ are listed in Table VI. Our results are nicely fit on Regge lines which justify authenticity of HQET formulation also.
     
   \begin{figure}[htp]
    \centering
    \includegraphics[width=9cm]{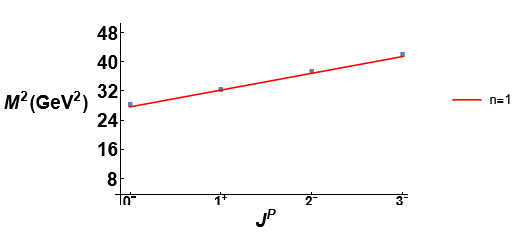}
    \caption{Regge trajectories for non strange bottom meson with unnatural parity in plane($M^2\rightarrow J^P$)}
    \label{fig:1}
\end{figure}
  \begin{figure}[htp]
    \centering
    \includegraphics[width=9cm]{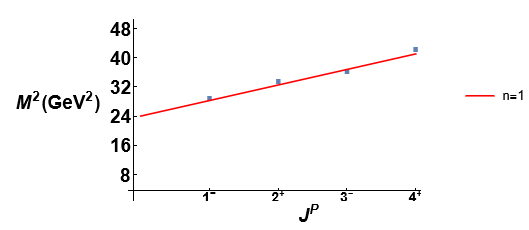} 
    \caption{Regge trajectories for non strange bottom meson with natural parity in plane($M^2\rightarrow J^P$)}
    \label{fig:2}
\end{figure} 
 \begin{figure}[htp]
    \centering
    \includegraphics[width=9cm]{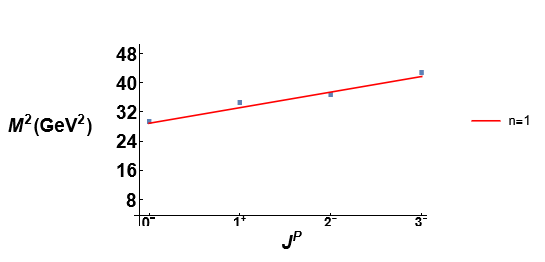}
    \caption{Regge trajectories for strange bottom meson with unnatural parity in plane($M^2\rightarrow J^P$)}
    \label{fig:3}
\end{figure}
\begin{figure}[htp!]
    \centering
    \includegraphics[width=9cm]{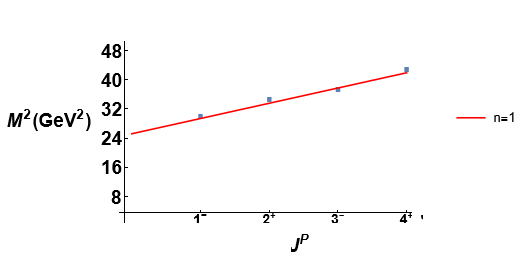}
    \caption{Regge trajectories for strange bottom meson with natural parity in plane($M^2\rightarrow J^P$)}
    \label{fig:4}
\end{figure}
\begin{table}[htp!]
    \centering
    \begin{tabular}{|c|c|c|c|}
    \hline
       Figure No.  & Slope($\alpha$)[MeV$^{-2}$] & Intercepts($\alpha_{0}$)\\
         \hline
         1 & 0.194717 & -5.30365\\
         \hline
         2 &  0.217415 & -5.07794\\
         \hline
         3 & 0.22186 & -6.52136\\
         \hline
         4 & 0.235764 & -5.90507\\
         \hline
        \end{tabular}
    \caption{Regge Slopes and Regge Intercepts }
    \label{tab:my_label60}
\end{table}
\section{Conclusion}
Heavy quark symmetry is important tool to describe spectroscopy of hadrons. Using available experimental as well as theoretical data on charm mesons and applying HQS, we computed masses for $1F$ bottom mesons spectra. With predicted masses of $1F$ bottom mesons, we analyzed decay widths for $1F$ with emission of pseudoscalar mesons and presents decay widths in form of coupling constants. These coupling constants are estimated on comparing our decay widths with  theoretical available total decay widths. The total decay widths may give upper bound on these coupling constants hence providing useful clue to other associated states of bottom mesons. Using our calculated bottom masses for $1F$, we construct Regge trajectories in ($n_r$, $M^2$ ) planes. Our predicted data nicely fit to them.  Our calculated masses and upper bounds findings may help experimentalists for looking into higher excited states.

\section{Acknowledgment}
The authors thankfully acknowledge the financial support by the
Department of Science and Technology (SERB/F/9119/2020), New Delhi.

\end{document}